\documentstyle[aps,multicol,epsfig]{revtex}
\begin{document}
\draft
\title{NMR Detection of Temperature-Dependent 
       Magnetic Inhomogeneities in URu$_{2}$Si$_{2}$}         
\author{O. O. Bernal}   
\address{Physics Department, California State University,
         Los Angeles, California 90032}
\author{B. Becker, J. A. Mydosh, G. J. Nieuwenhuys, A. A. Menovsky, 
        P. M. Paulus and H. B. Brom}
\address{Kamerlingh Onnes Lab, 2300 RA Leiden, The Netherlands}
\author{D.E.~MacLaughlin}
\address{University of California, Riverside, California 92521}
\author{H.G.~Lukefahr}   
\address{Whittier College, Whittier, California 90608}
\date{June 30, 1999}
\maketitle

\begin{abstract}
We present $^{29}$Si-NMR relaxation and spectral data in URu$_2$Si$_2$.
Our echo-decay experiments detect slowly fluctuating magnetic field 
gradients. 
In addition, we find that the echo-decay shape (time dependence) varies 
with temperature $T$ and its rate behaves critically near $T_{\rm N}$, 
indicating a correlation between the gradient fluctuations and the transition 
to small-moment order. 
$T$-dependent broadening contributions become visible 
below ${\sim}$100~K and saturate somewhat above 
$T_{\rm N}$, remaining saturated at lower temperatures.
Together, the line width and shift suggest partial lattice distortions 
below $T_{\rm N}$.
We propose an intrinsic minority phase below $T_{\rm N}$ and compare our 
results with one of the~current~theoretical~models.\vspace{3mm}
\end{abstract}

\pacs{PACS numbers: 71.27.+a, 76.60.-k, 75.10.-b, 75.25.+z, 75.30.-m}

\begin{multicols}{2}
\settowidth{\columnwidth}{aaaaaaaaaaaaaaaaaaaaaaaaaaaaaaaaaaaaaaaaaaaaaaaaa}

In URu$_{2}$Si$_{2}$, there is coexistence of magnetic order 
($\mu\sim{0.04}\mu_{B}$/U, $T_{\rm N}\sim{17.5}$~K) and 
superconductivity ($T_{\rm c}\sim{1.2}$~K)~\cite{first}.
Much attention has been focused on the transition at 
$T_{\rm N}$~\cite{first,santini,mydosh},
with studies ultimately suggesting a quadrupolar order 
parameter~\cite{santini}. 
However, no direct evidence has been found for quadrupolar 
ordering.
URu$_{2}$Si$_{2}$ has little or no residual chemical disorder
or magnetic frustration~\cite{mydosh}, but evidence for 
chemical order does not guarantee magnetic homogeneity~\cite{oob}.
Here, we report $^{29}$Si-NMR data 
(line width~$\sigma$, shift~$K$,  and spin-echo decay rate $R$) 
which show $T$-dependent magnetic inhomogeneities below
${\sim}$100~K that correlate with the unusual magnetic order.

\begin{figure}[htb]
\begin{center}
\leavevmode
\epsfig{figure=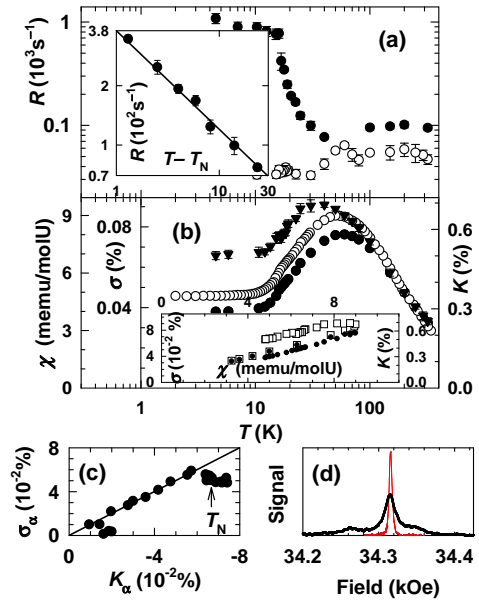,height=8.5cm}
\end{center}
\caption{(a) $R$ vs. $T$ ($H{\|}c$-axis=9.4~T); 
         circles: HE (solid), CP (open).
         Inset: $R_{\rm HE}$ vs. $T-T_{\rm N}$; $T_{\rm N}{\sim}17$~K 
         (line: fit to 
         $AT_{\rm N}^{1/2}/(T-T_{\rm N})^{1/2}$; $A{\sim}123$~s$^{-1}$).
         (b) Circles: $\chi$ (open) and $K$ (solid) vs. $T$.
         Triangles: $\sigma$ vs.~$T$.
         Inset: $\sigma$ (squares) and $K$ (circles)
         vs. $\chi$, $T$ implicit.
         (c)~$\sigma_\alpha(T)$~vs.~$K_\alpha(T)$. 
         (d) Spectra ($H{\bot}c$-axis): $T=25$~K (narrower line),
         4.2~K (broader line).}
\label{fig:1}
\end{figure}

The sample was an oriented (alignment$\agt$~95\%), epoxy-embedded
powder (particle size$\alt$~100~$\mu$m). 
Fig.~\ref{fig:1}(a) shows $R(T)$ obtained using conventional 
Hahn-echo (HE) [$R_{\rm HE}$: solid circles] and Carr-Purcell 
(CP) [$R_{\rm CP}$: open circles] sequences~\cite{nmr} 
($H\|{c}$-axis=9.4~T).
$R_{\rm CP}$ is nearly $T$-independent, whereas, $R_{\rm HE}$ is not.
Also, $R_{\rm HE}{\gg}R_{\rm CP}$ for $T\alt$40~K\@. 
The shape of the HE-decay signal as a function of time
$t$ (2$\times$pulse-separation) can be fitted to 
$\exp{-(R_{HE}t)^2}$ or $\exp{-(R_{HE}t)^3}$ for $T>40$~K,
$\exp{-(R_{HE}t)}$ for 40~K$>T>T_{\rm N}$, and
$\exp{-(R_{HE}t)^{1/2}}$ for $T<T_{\rm N}$.
$R_{HE}$ remains close to its maximum value for $T<T_{\rm N}$ 
with no indication of dying out at lower $T$'s.
Crude estimates of the homogeneous linewidth in the dilute
limit~\cite{nmr}
(natural abundances of $^{29}$Si, $^{101,99}$Ru: 5\%, 17\%, 13\%)
come out too low ($\sim$20\% of $R_{\rm HE}$ at low $T$) and appear 
inconsistent with the observed $T$-dependent shape of the HE decay.
Our $R$ measurements are at odds with previously reported
experiments~\cite{kohori}.

The NMR spectra was fit to a single gaussian function.
$\sigma(T)$, $K(T)$ (from the fit), and the magnetic susceptibility
$\chi(T)$ (from a SQUID magnetometer) are presented in Fig.~\ref{fig:1}(b).
For $T\agt$100~K, both $K$ (solid circles) and $\sigma$ (triangles) 
display the same $T$-dependence as $\chi$ (open circles).
For $T\alt$100~K, there are non-linearities in the $K$
and $\sigma$ vs. $\chi$  plots (inset).
The excess width is inhomogeneous,
i.e., $R_{\rm HE}$ can only account for about 1\% of the total.

In a solid, there is no reason to expect a large difference 
between $R_{\rm HE}$ and $R_{\rm CP}$ (Fig.~\ref{fig:1}(a), $T\alt$40~K).
In liquids it is quite common to find such difference, because
the combination nuclear diffusion/magnetic field gradient creates 
more dephasing of the echo signal in the HE method; the CP method
measures a smaller $R$ by eliminating the dephasing effect~\cite{nmr}.
By comparison, having $R_{\rm HE}{\gg}R_{\rm CP}$ ($T\alt$40~K)
here implies randomly moving or fluctuating modulations 
of the U-spin system that can be sensed by the static $^{29}$Si-nuclei.
We believe that this is the first time that such effect is 
measured in a heavy-fermion system.

The $T$-dependence of $R_{\rm HE}$ and $\sigma$ could be suggesting
local charge-density wave (CDW) structures as a source of U-spin
modulations.
In CDW systems, $R$ can display a $(T-T_{0})^{-1/2}$ dependence for 
$T{\agt}T_{0}$ ($T_{0}$: CDW transition temperature)~\cite{cdw}
[inset Fig.~\ref{fig:1}(a); $T_{0}{\rightleftharpoons}T_{N}$].
Also, incommensuration/randomness of forming CDW's
would distribute the U-$^{29}$Si transferred hyperfine coupling 
locally and explain the observed $T$-dependence of~$\sigma$.
In addition, there is no clear change in NMR line shape below $T_{\rm N}$, 
so only a fraction of the $^{29}$Si nuclei must be affected;
the majority would sample an underlying paramagnetic fluid.
Given the coexistence of magnetism and 
superconductivity~\cite{first} and that two kinds of transitions 
seem to be needed to explain the macroscopic anomalies~\cite{santini},
an interpretation in terms of two different phases 
(as opposed to introducing a $T$-dependent hyperfine coupling)
seems plausible. 
Therefore, we fit the data using the simple 
relations $K=K_0+a\chi(T)+K_\alpha(T)$, and
$\sigma^2=\sigma_0^2+({\delta}a)^2\chi^2(T)+\sigma_\alpha^2(T)$.
Here $K_0=0.06$\% (contact hyperfine interaction),
$a=3.5$~kOe/$\mu_B$,
$\sigma_0=0.005$\%, and $({\delta}a)\sim$0.5~kOe/$\mu_B$.
$K_\alpha(T)$ and $\sigma_\alpha(T)$ are the contributions 
due to the second phase.

In Fig.~\ref{fig:1}(c), a plot of $\sigma_\alpha$ vs. $K_\alpha$ shows 
proportionality at high $T$ (slope$\sim{-1}$) and a 
breakdown of linearity somewhat above $T_{\rm N}$.
This is more qualitative evidence that the inhomogeneous 
magnetism is correlated with the unusual transition.
The slope in this plot represents the fractional width of the 
distribution of U-$^{29}$Si transferred hyperfine couplings
for the second phase.
The abrupt slope change just above $T_{\rm N}$ 
(from~${\sim}-1$ to~${\sim}0$) would signal an intrinsic rearrangement
of the U-spins in this phase with respect to the $^{29}$Si nuclei, 
i.e., partial lattice distortions, since no distortion affecting the 
whole crystal has been detected (see~\cite{first,santini}). 
The saturation of $\sigma_\alpha$ below $T_{\rm N}$ indicates that 
these distortions remain~at~lower~$T$.

In conclusion, we have found evidence for random or incommensurate
structures in URu$_{2}$Si$_{2}$ at low $T$. 
We propose slowly moving and/or fluctuating charge-density modulations,
which could in turn serve as domain walls to antiferromagnetic regions
consistent with the resistance anomaly measured at 
$T_{\rm N}$~\cite{first}.
One could also argue that our results
(local distortions near, but not at, $T_{\rm N}$, as well as
U-spin modulation fluctuations when $T{\rightarrow}T_{\rm N}$)
are consistent with the picture presented in Ref.~\cite{santini}.
Finally, the spectra for $H\bot{c}$-axis are also consistent with U-spins 
modulations~\cite{santini}, Fig.~\ref{fig:1}(d).
Systematic field and orientation dependence studies of the NMR 
anomaly are being carried out to make progress in this~direction.

We acknowledge helpful discussions with R.E.~Walstedt and D.L.~Cox.
This research was supported by an award from Research 
Corporation and NSF grant DMR-9820631.

\end{multicols}

\end{document}